\documentclass[conference]{IEEEtran}

\usepackage{xcolor}
\usepackage{setspace}
\usepackage{graphicx}
\usepackage{lipsum}
\usepackage{svg}
\usepackage{amsmath,amssymb,amsfonts}
\usepackage{algorithmic}
\usepackage{tabularray}
\usepackage{stfloats}
\usepackage{paralist}
\usepackage{subcaption}
\usepackage{cite}
\usepackage{textcomp}
\usepackage{comment}
\usepackage[hidelinks]{hyperref}
\usepackage{sourcecodepro}
\usepackage[T1]{fontenc}

\usepackage{pifont}

\usepackage{listings}
\definecolor{codegreen}{rgb}{0.2,0.5,0.2}
\definecolor{codegray}{rgb}{0.5,0.5,0.5}
\definecolor{codepurple}{rgb}{0.58,0,0.82}
\lstdefinestyle{mystyle}{
    backgroundcolor=\color{white},   
    commentstyle=\color{codegreen},
    keywordstyle=\color{magenta},
    numberstyle=\tiny\color{codegray},
    stringstyle=\color{codepurple},
    basicstyle=\ttfamily\footnotesize,
    breakatwhitespace=false,         
    breaklines=true,                 
    captionpos=b,                    
    keepspaces=true,                 
    numbers=left,                    
    numbersep=5pt,                  
    showspaces=false,                
    showstringspaces=false,
    showtabs=false,                  
    tabsize=2
}
\gdef\DefinePerson#1#2{
    \expandafter\gdef\csname #1@color\endcsname{#2}
    
    \expandafter\newcommand\csname #1\endcsname[1]{%
        \textcolor{\csname #1@color\endcsname}{\textbf{#1} ##1}%
    }
}

\DefinePerson{Julien}{orange}
\DefinePerson{Paolo}{blue}
\DefinePerson{Mohamed}{purple}
\DefinePerson{Cem}{green}

\def\BibTeX{{\rm B\kern-.05em{\sc i\kern-.025em b}\kern-.08em
    T\kern-.1667em\lower.7ex\hbox{E}\kern-.125emX}}

\begin{document}


\title{Bombyx: OpenCilk Compilation for FPGA Hardware Acceleration}


\author{
    \IEEEauthorblockN{ %
        Mohamed Shahawy,
        Julien de Castelnau,
        Paolo Ienne
    }
    \IEEEauthorblockA{
    École Polytechnique Fédérale de Lausanne, CH-1015 Lausanne, Switzerland
    \\ 
    \{mohamed.shahawy, julien.decastelnau, paolo.ienne\}@epfl.ch}
}

\maketitle

\begin{abstract}
\emph{Task-level parallelism} (TLP) is a widely used approach in software where independent tasks are dynamically created and scheduled at runtime.
Recent systems have explored architectural support for TLP on \emph{field-programmable gate arrays} (FPGAs), often leveraging high-level synthesis (HLS) to create \emph{processing elements} (PEs).
In this paper, we present \emph{Bombyx}, a compiler toolchain that lowers OpenCilk programs into a Cilk-1-inspired intermediate representation, enabling efficient mapping of CPU-oriented TLP applications to spatial architectures on FPGAs.
Unlike OpenCilk’s implicit task model, which requires costly context switching in hardware, Cilk-1 adopts explicit continuation-passing---a model that better aligns with the streaming nature of FPGAs.
Bombyx supports multiple compilation targets: one is an OpenCilk-compatible runtime for executing Cilk-1-style code using the OpenCilk backend, and another is a synthesizable PE generator designed for HLS tools like Vitis HLS.
Additionally, we introduce a decoupled access-execute optimization that enables automatic generation of high-performance PEs, improving memory-compute overlap and overall throughput.

\end{abstract}


\section{Introduction}
\label{sec:intro}


\emph{Task-level parallelism} (TLP) is a technique utilized to exploit parallelism within control-dominated algorithms, where parallelism emerges dynamically during execution. 
TLP involves a set of functions, termed \emph{tasks}, capable of spawning each other and themselves during execution.
Various software frameworks support TLP, such as Intel Thread Building Blocks~\cite{tbb_2019}, Cilk Plus, and OpenCilk~\cite{schardl:OpenCilk:2023}. 
OpenCilk represents the most recent framework, demonstrating superior performance compared to its counterparts. 
Notably, multiple hardware frameworks, including ParallelXL~\cite{chen:parallelxl:2018} and HardCilk~\cite{shahawy:hardcilk:2024}, have been designed to facilitate TLP on \emph{field-programmable gate arrays} (FPGAs) platforms. 
These frameworks automate generating a hardware TLP scheduler for a given application.

Despite the availability of hardware frameworks supporting TLP on FPGAs, there has been a lack of automation for extracting \emph{processing elements} (PEs) code from software frameworks and directly compiling it using \emph{high-level synthesis} (HLS). 
Typically, TLP hardware frameworks rely on users to supply the requisite code for PEs. 

TLP is represented in software using two approaches.
Firstly, the fork-join approach used by OpenCilk (see Figure~\ref{lst:fib_oc}) where \emph{spawn} and \emph{sync} are analogous to fork and join, respectively. 
We define this approach as the \emph{implicit} approach, not because \emph{sync} itself is subtle, it clearly acts as a barrier, but because it implicitly entails suspending execution until all spawned functions complete.
Such \emph{implicit} approach requires suspending the function to execute the spawned functions, where context data required after the barrier is saved and retrieved only after the \emph{sync} barrier---i.e, context switching.
%
%
Implementing a PE in hardware capable of context switching is not straightforward because current HLS tools do not support it, and it would require significant RTL engineering, as it requires saving the whole state of the circuit.

The other approach is the \emph{explicit continuation passing} approach used by the earlier Cilk-1~\cite{Blumofe:cilk:1995} software scheduler, see Figure~\ref{lst:fib_cps}.
The two approaches are largely equivalent, and it was proven that any \emph{implicit} program can be converted to the \emph{explicit} form~\cite{JoergJan96}.
The \emph{explicit} form replaces the \emph{sync} statement with two new keywords: \emph{send\_argument} and \emph{spawn\_next}.
These keywords are explained later in detail; however, the main takeaway is that in the explicit form, functions are fissioned around the \emph{sync} point to terminating functions---those that execute atomically rather than requiring synchronization and context switching.











\begin{figure}[tb]
    \centering
    \begin{tabular}{c}
        \lstset{numbers=left, showspaces=false, showstringspaces=false, tabsize=2, breaklines=true}
\begin{lstlisting}[language=c++, basicstyle=\footnotesize\ttfamily]
int fib(int n) {
    if (n < 2)
        return n;
    int x = cilk_spawn fib(n-1);
    int y = cilk_spawn fib(n-2);
    cilk_sync;
    return x + y;
}
\end{lstlisting}
    \end{tabular}
    \caption{OpenCilk code for a program computing Fibonacci. The use of \emph{spawn} and \emph{sync} represents implicit parallelism---that is, the programmer guarantees by the \emph{spawn} keywords that the invoked function is independent, and the keyword \emph{sync} implicitly synchronizes the spawned functions. }
    \vskip-0.15in
    \label{lst:fib_oc}
\end{figure}


%
%
%
%



The \emph{explicit} approach for TLP is more natural to use when targeting hardware. However, due to requiring the fission of code into terminating functions, it is less intuitive and convenient to write software in. Thus, it did not survive within the Cilk family, and OpenCilk today adopts the implicit model.
To provide an automatic flow from software TLP to Hardware TLP, we introduce \emph{Bombyx}, a tool designed to automate the extraction and compilation of functional code from fork-join software frameworks like OpenCilk into an \emph{explicit-style} \emph{intermediate representation} (IR). 
Bombyx supports multiple compilation targets: producing C++ code compatible with Vitis HLS for generating PEs for the open-source HardCilk~\cite{shahawy:hardcilk:2024} FPGA system, and generating executable Cilk-1-like code using the OpenCilk runtime environment for verification.

\begin{figure}[tb]
    \centering
    \begin{tabular}{c}
        \lstset{numbers=left, showspaces=false, showstringspaces=false, tabsize=2, breaklines=true}
\begin{lstlisting}[language=c++, basicstyle=\footnotesize\ttfamily]
task fib (cont int k, int n) {
    if (n < 2) {
        send_argument(k, n);
    } else {
        cont int x,y;
        spawn_next sum(k, ?x, ?y);
        spawn fib(x, n-1);
        spawn fib(y, n-2);
    }
}
task sum (cont int k, int x, int y) {
    send_argument(k, x + y);
}
\end{lstlisting}
    \end{tabular}
    \caption{Cilk-1 code for computing Fibonacci. Replacing the \emph{sync} keyword, compared to Figure~\ref{lst:fib_oc}, with \emph{send\_argument} and \emph{spawn\_next} represents explicit parallelism---that is, the user explicitly specifies the function to be executed after all the spawned functions complete, as opposed to resuming within the same function after a sync.}
    \vskip-0.15in
    \label{lst:fib_cps}
\end{figure}

This paper is structured as follows. 
Section~\ref{sec:compiler-flow} discusses the design of Bombyx, starting with the structure of the IR and how an OpenCilk program is translated into it. 
In Section~\ref{sec:hardcilk}, the code generation targeting the HardCilk FPGA framework is presented. 
Finally, Section~\ref{sec:dae} explains how Bombyx can aid programmers in applying decoupled access-execute optimizations on TLP code. We briefly evaluate the impact of these optimizations in our evaluation, in Section~\ref{sec:eval}.




\section{Bombyx}
\label{sec:bombyx}

\begin{figure}[tb]
\centering
\includegraphics[width=0.48\textwidth]{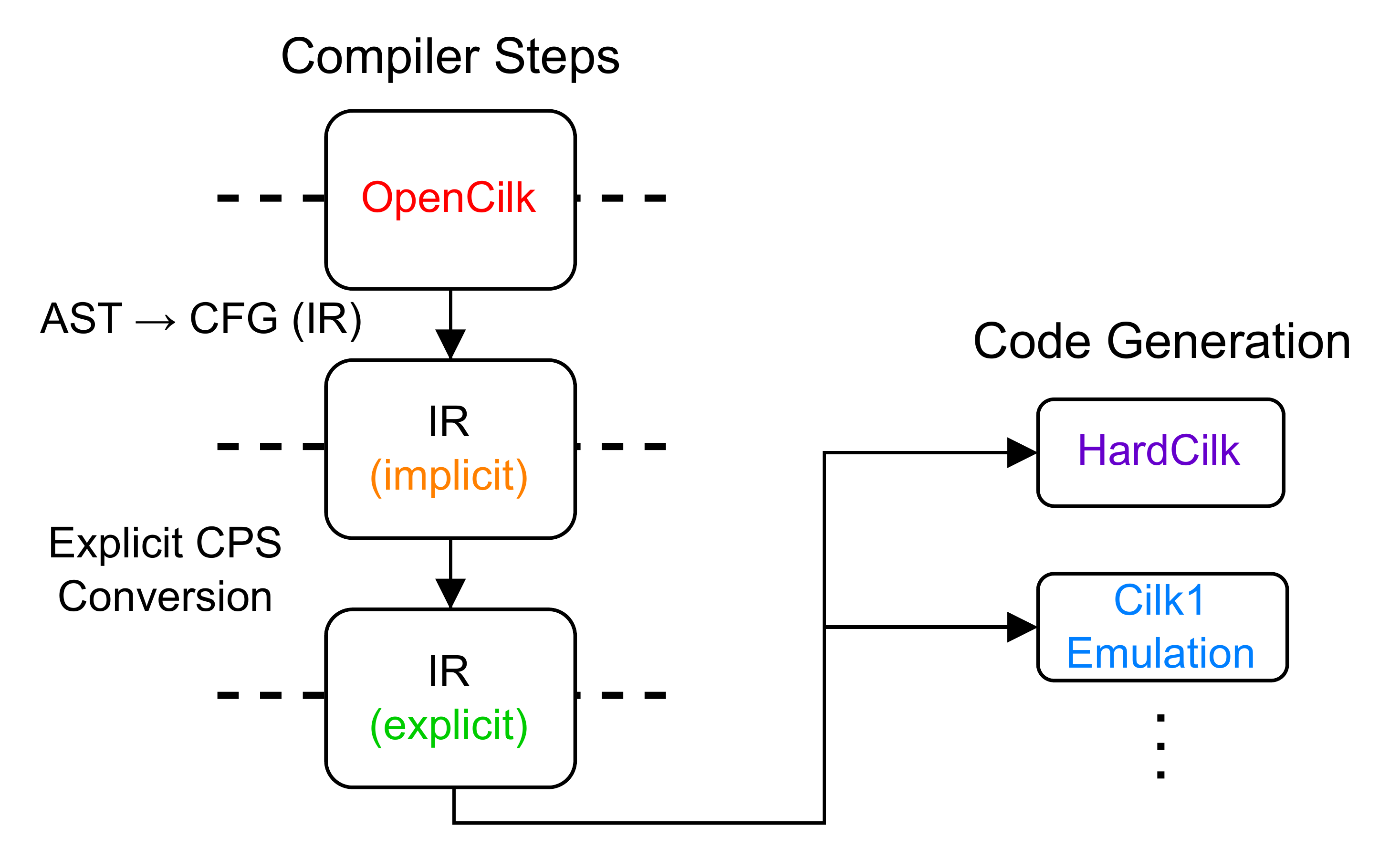}
\caption{
Compilation flow of Bombyx for an OpenCilk TLP program: (1) OpenCilk Clang extracts the \emph{abstract syntax tree} (AST); (2) the AST is converted into an implicit \emph{intermediate representation} (IR), where optimizations are applied; (3) the implicit IR is lowered to an explicit IR, which maps to hardware task-parallel frameworks.
}
\vskip-0.15in
\label{fig:flow}
\end{figure}
\vskip-0.5pt

This section explains the compilation flow of Bombyx, depicted graphically in Figure~\ref{fig:flow}. 
The input to the tool is a C/C++ OpenCilk program targeting CPUs.
This input is converted to an \emph{abstract syntax tree} (AST) using the front end of the OpenCilk version of \emph{Clang}.
Bombyx converts the AST into an \emph{intermediate representation} (IR), known as the \emph{implicit IR}.
Bombyx avoids lowering the AST to OpenCilk's \emph{TAPIR}, Figure~\ref{fig:IRs}~(a), as the latter uses lower-level compiler constructs that make it harder to generate C++ code for HLS as close as possible to the original implicit code. 
The DAE optimization is applied at this stage, before transforming to the \emph{explicit IR}---representing \emph{explicit} TLP.
It is easy to directly lower the \emph{explicit IR} to PEs that target hardware TLP. 
We follow the example in Figure~\ref{lst:fib_oc} through the compilation process.



Converting \emph{implicit} TLP to \emph{explicit} TLP requires identifying the dependencies across the \emph{sync} barrier.
Such dependencies identify the program state that needs to be explicitly recorded in memory using the \emph{spawn\_next} keyword. 
\emph{spawn\_next} creates a task in memory waiting for the dependencies before execution, known as the \emph{continuation} task.
The task is represented as a data structure known as the \emph{closure}, which includes the ready arguments, placeholders for the anticipated dependencies, and a pointer that the task shall return to.
The \emph{send\_argument} keyword replaces the implicit \emph{return} by explicitly writing arguments into placeholders within waiting closures created by \emph{spawn\_next}, and notifying the scheduler accordingly.

%
%
%
%
%

\subsection{From OpenCilk to Explicit Intermediate Representation}
\label{sec:compiler-flow}

The first step of the flow is to convert the AST of the program, generated by OpenCilk \emph{Clang}, to an IR where optimizations and lowering steps are easier to perform.
The Bombyx IR is a \emph{control-flow graph} (CFG), which is a directed acyclic graph of \emph{basic blocks}, each containing a sequence of C statements where the program control flow is through the edges connecting blocks.
Basic blocks are \emph{terminated} by statements that affect the control flow, such as \emph{if}, \emph{for}, \emph{return}, etc. We additionally consider \emph{sync} to affect the control flow because we consider it as a function terminator in the \emph{explicit IR} that we later discuss.

\begin{figure*}
\includegraphics[width=\textwidth]{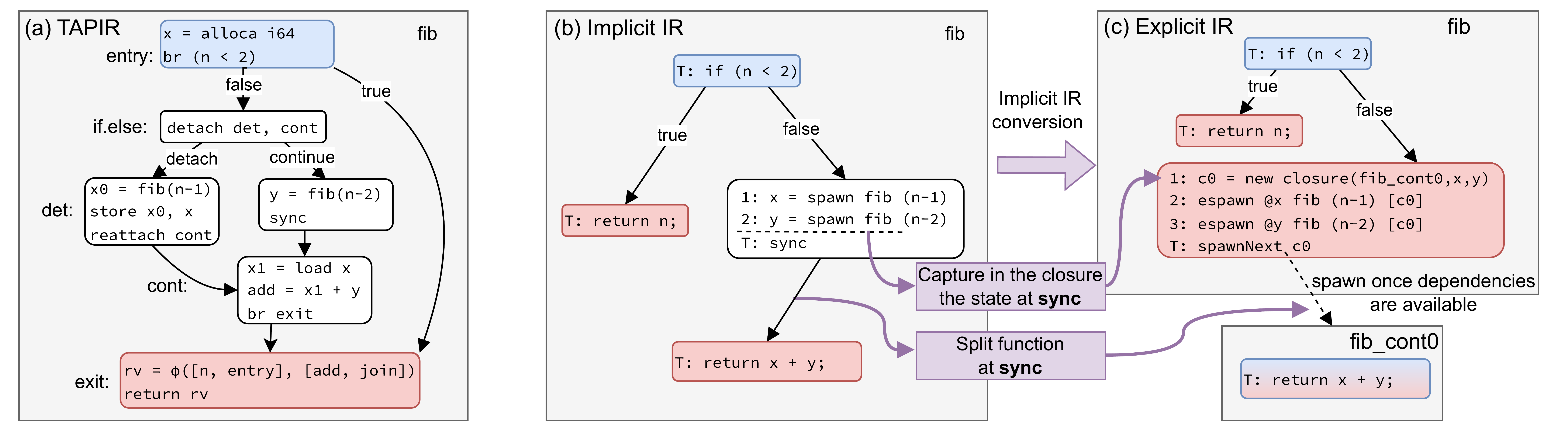}
\caption{Intermediate representations. Blue and red blocks are entry and exit blocks, respectively. In the Bombyx IRs, T: denotes the terminating statement of the basic block. (a) \emph{TAPIR} is an LLVM derivative used by OpenCilk for compilation. This IR changes the structure of the original C++ code to apply optimizations for software, which makes it harder to generate readable explicit C++ code for HLS. (b) The \emph{implicit IR}, an intermediate IR used by Bombyx as it easier to convert to \emph{explicit IR} than using the AST generated by the OpenCilk Clang frontend. This IR preserves the original structure of the C++ code. (c) The \emph{explicit IR}, the final IR generated by Bombyx, which can be directly mapped to hardware primitives.
}
\vskip-0.10in
\label{fig:IRs}
\end{figure*}

\textbf{The implicit IR.}
The primary purpose of this IR is to serve as the step between the AST and the explicit code, as the explicit conversion is much easier performed on a control flow graph rather than an AST, where aspects like scoping make the process more difficult. 
%
%
%
Each function in the input program has its own control flow graph, with exactly one \emph{entry block} and one or more \emph{exit blocks}. 
The entry block has no incoming edges, while an exit block is any block without any outgoing edges. An example CFG for the Fibonacci program of Figure~\ref{lst:fib_oc} is shown in Figure \ref{fig:IRs}~(b). 
We call this CFG the \emph{implicit IR}, as the \emph{sync} statements of OpenCilk have not yet been converted to explicit continuations that are usable for FPGA synthesis.

\textbf{The explicit IR.}
Given a control flow graph of an \emph{implicit IR} (exemplified in Figure~\ref{fig:IRs}~(b)), we explain the conversion procedure to \emph{explicit IR}, as in Figure~\ref{fig:IRs}~(c).
Bombyx traverses the control flow graph and partitions it into a set of subgraphs, referred to as \emph{paths}.
A new path is initiated upon encountering either a function-terminating basic block (e.g., a \emph{return} statement) or a block containing a \emph{sync} operation. 
Each path constitutes a self contained terminating function, though inter path dependencies may persist and must be accounted for during subsequent analysis.

After identifying the paths, two represented by the greyed out zones in Figure~\ref{fig:IRs}~(c), Bombyx links the separated functions in the \emph{explicit IR} using the Cilk-1 keywords to keep the dependencies expressed by the \emph{implicit IR}.
At the boundaries between paths, \emph{sync} is replaced with a \emph{spawn\_next} call to the separated function whose entry block corresponds to the successor of the boundary block. 
Alongside \emph{spawn\_next}, Bombyx inserts a \emph{closure declaration} at the block containing the \emph{spawn} calls. 
These closures contain any ready arguments, placeholders for the anticipated dependencies, and a return pointer.
Finally, to handle data passing between the subgraphs, spawns are converted from a value-returning expression (like in a function call) to a task spawn whose explicit return address is set to the corresponding field of the closure. 
Return statements are retained in the \emph{explicit IR} as they can be converted to the explicit \emph{send\_argument} when required.

\subsection{Lowering to HardCilk}
\label{sec:hardcilk}
Currently, Bombyx supports lowering of the \emph{explicit IR} to two backends. 
The first is HardCilk, a system for TLP on FPGAs. 
The second is a Cilk-1 ``emulation layer'' that implements each of the Cilk-1 constructs (spawn, spawn\_next, send\_argument) in OpenCilk itself. 
We omit discussion of this backend here, as its primary purpose is to verify the equivalence of the original program in software once compiled.

HardCilk is an open-source architecture generator for hardware accelerators leveraging TLP. 
It provides a customizable hardware \textit{work-stealing scheduler} to load balance task-parallel PEs, which could be implemented in C++ using HLS. 
PEs access the scheduler by a set of stream interfaces, effectively implementing \emph{spawn}, \emph{spawn\_next}, and \emph{send\_argument}.

Mapping the \emph{explicit IR} directly to C++ HLS PEs is straightforward; however, there are a number of additional requirements to automatically create HardCilk accelerators. 
HardCilk PEs are tedious to write by hand for a number of reasons that we automate through the HardCilk lowering backend. 
First, each task closure, representing the function parameters, needs to be aligned to a certain size (128, 256 bits, etc.), to be easily implementable in hardware. 
Without Bombyx, padding is added manually to compensate.
Second, HardCilk uses a \emph{write buffer} module when issuing a \emph{spawn\_next} or a \emph{send\_argument} to free the PE for further processing.
The write buffer requires the HLS code to include extra information about the argument/task being written; adding such information manually can be prone to errors and is automated by Bombyx.
Finally, HardCilk requires a JSON configuration file serving as a descriptor for the relations among tasks in the system.
The JSON contains the size of closures in the system, a list of which tasks a given task may \emph{spawn}, \emph{spawn\_next}, or \emph{send\_argument} to, and others. 
%
%
These transformations are performed using static analysis on lowering to HardCilk.


%


\subsection{Decoupled access execute}
\label{sec:dae}

A common optimization in hardware is that of \emph{decoupled access-execute} (DAE).
In a naive PE implementation, the same unit issues memory requests and performs computation. 
The result is that a memory-related stall causes the PE to wait idle, preventing it from doing useful work and decreasing overall throughput. 
Pipelining is a common mitigation supported by HLS tools, where the memory access and execute portions of the PE's code are broken into pipelined stages, where useful computation can be completed while memory is stalled.

However, pipelining is limited in industrial statically scheduled HLS tools, like Vitis. 
When the latency of operations in the PE cannot be determined statically, for example, a loop with a data dependent bound, the tool cannot fully pipeline the computation. 
Figure~\ref{lst:dae_benchmark} shows an example of such a program. 
The code is expressed here in OpenCilk, but it looks similar once mapped to C++ targeting HLS. 
The \emph{visit} routine performs a parallel breadth-first traversal of a graph starting at a node $n$. 
In this program, the memory access on line 3 stalls the remainder of the computation, which visits nodes recursively based on the adjacency list. 
Pipelining the unit should allow accesses for future iterations to be made while new tasks are spawning, but in Vitis HLS and similar tools, it fails due to the variable latency of the for-loop on line 7. 
That is, it is not possible to overlap the stages because the HLS tool cannot determine when to schedule the memory access; the latency of the for loop stage depends on that access.

\begin{figure}[tb]
    \centering
    \begin{tabular}{c}
        \lstset{numbers=left, showspaces=false, showstringspaces=false, tabsize=2, breaklines=true}
\begin{lstlisting}[language=c++, basicstyle=\footnotesize\ttfamily]
void visit(struct graph_t *g, struct node_t *n) {
    // Access to be decoupled
    struct node_t nd = *n; 
    if (nd.visited)
        return;
    n->visited = true;
    for (int i = 0; i < nd.len; i += 1)
        cilk_spawn visit(g, 
            &g->nodes[nd.neighbors[i]]);
}
\end{lstlisting}
    \end{tabular}
    \caption{
    Parallel breadth-first graph traversal using OpenCilk.
    }
    \vskip-0.15in
    \label{lst:dae_benchmark}
\end{figure}


To avoid these limitations, under TLP, it is possible to express the access and execute stages as \textit{different tasks}. 
This removes the need for the HLS tool to determine how to pipeline the stages: the PEs work completely independently, and the task scheduler orchestrates their execution. 
In this way, they work elastically: the PEs start tasks whenever they have data, instead of being artificially bottlenecked by a conservative schedule created by the HLS tool. 
We note that software compilers do not apply such optimizations, as CPUs have caches that reduce access latency and render such optimization useless for software.

Bombyx supports DAE as an optimization, although for now the transformation is guided through manual insertion of pragmas in the code. 
The programmer can insert the pragma in their source OpenCilk code at the point where the memory access occurs, and the compiler will split that operation and the code after it into separate tasks. 
Behind the scenes, the pragma prompts the compiler to extract the line below it into its own function, and replace that line of code with a spawn to that function, followed by a sync. 
Once converted to explicit style, the result is that at the original point of the memory access, a new task for that access is spawned, and it is passed a continuation to the task for the code after it, on which \emph{spawn\_next} is invoked. 

\section{Evaluation}
\label{sec:eval}

%

%

%

%

%


%

%



%


We evaluate the effectiveness of the DAE optimization supported by Bombyx, using the open-source HardCilk framework.
The benchmark is the previously discussed graph traversal program in Figure~\ref{lst:dae_benchmark}. 
A structure representing the adjacency list for the passed node is first loaded, then the node is marked as visited, and, finally, the adjacent nodes are visited recursively in parallel. 
We insert the DAE optimization pragma \verb|#PRAGMA BOMBYX DAE| supported by Bombyx on line 2 to separate the memory access for the adjacency list into its own access task---the rest of the code becomes an execute unit. 

We use the HardCilk backend of Bombyx to compare the accelerator generated with and without the DAE pragma applied. 
The execution time required to traverse an entire graph is used as a performance measure. 
We use two graphs as a dataset, each synthetically generated as a tree with depths $D=7$ and $9$, and branch factor $B=4$ for each node. 
In total, the graphs are of size $\frac{B^D-1}{B-1} = 5,461$ and $87,381$ respectively. 
%
%
We generate two HardCilk systems using the authors' open-source generator.
We specify the number of PEs as one in the non-DAE case, and one for each type of task in the DAE case.
%

%

\begin{figure}
\centering
\begin{tabular}{|c|c c c|}
\hline & LUT & FF & BRAM \\ \hline
\textbf{Non-DAE} & 2657 & 2305 & 2 \\ \hline
Spawner & 133 & 387 & 0 \\
Executor & 1999 & 1913 & 2 \\
Access & 1764 & 1164 & 2 \\
\textbf{DAE (total)} & 3896 & 3464 & 4 \\ \hline
\end{tabular}
\caption{Synthesis results for DAE optimization PEs.}
\vskip-0.15in
\label{tab:synth}
\end{figure}

Overall, we observe a 26.5\% reduction in runtime comparing the DAE and non-DAE versions.
We also test the resource utilization incurred by the DAE optimization PEs.
For synthesis, we use Vivado 2024.1 on a Xilinx Alveo U55C with part number \verb|xcu55c-fsvh2892-2L-e|, targeting a frequency of 300~MHz.
Figure \ref{tab:synth} displays the synthesis results for the PEs of each version. 
Overall, applying DAE results in an 47\% increase in LUTs and 50\% in FFs. 
In particular, the size of the spawner and executor PEs combined is nearly equal to the non-DAE size. 
For the access task, an RTL implementation of a single data-parallel PE would benefit here, as it amortizes its cost among all executors.
%
%
In the future, Bombyx could integrate such an implementation as a black-box primitive invoked whenever the DAE pragma is applied, rather than using HLS.



\section{Conclusion}
We present Bombyx, a compiler designed to aid in porting task-parallel software to bespoke hardware frameworks. 
Bombyx compiles from OpenCilk to an IR which uses \emph{continuation passing style}, an ideal form for expressing TLP on FPGAs. 
Currently, it can generate PEs in C++ HLS for the HardCilk framework, but its design makes it extensible to other targets. 
Bombyx can also apply decoupled-access execute optimizations; evaluated on a synthetic graph traversal benchmark, it decreases runtime by up to 26.5\%.

\bibliographystyle{IEEEtran}
\bibliography{bib/memory}

\end{document}